\documentclass[floatfix,twocolumn,prl,showpacs]{revtex4}
\usepackage{graphicx}
\usepackage{subfigure}
\usepackage{amssymb,amsmath}

\newcommand{\coloneq}{\; \colon \mspace{-12.0mu} =}

\begin{document}

\title{Scattering of low Reynolds number swimmers}

\author{G. P. Alexander, C. M. Pooley, and J. M. Yeomans}

\affiliation{Rudolf Peierls Centre for Theoretical Physics, 1 Keble Road, Oxford, OX1 3NP, United Kingdom.}

\date{\today}

\pacs{47.63.mf, 47.63.Gd, 47.15.G-}


\begin{abstract}
We describe the consequences of time reversal invariance of the Stokes' equations for the hydrodynamic scattering of two low Reynolds number swimmers. For swimmers that are related to each other by a time reversal transformation this leads to the striking result that the angle between the two swimmers is preserved by the scattering. The result is illustrated for the particular case of a linked-sphere model swimmer. For more general pairs of swimmers, not related to each other by time reversal, we find hydrodynamic scattering can alter the angle between their trajectories by several tens of degrees. For two identical contractile swimmers this can lead to the formation of a bound state.
\end{abstract}
\maketitle

The motile behaviour of micron sized organisms offers an insight into a physical environment very different to our own. Micron length scales correspond to low Reynolds number conditions where viscous forces dominate over the effects of inertia. Since Taylor's seminal paper~\cite{taylor} there has been considerable progress in our understanding of how low Reynolds number swimmers generate their motility~\cite{lighthill76,purcell77,shapere89a,stone96,wiggins98}. In the past few years this has included both the development of artificial microswimmers~\cite{dreyfus05} and a number of simple theoretical models~\cite{ramin,avron05}. 

A topic of growing interest is the role played by hydrodynamic interactions in determining low Reynolds number swimming. These interactions may be expected to be substantial because of the long range nature of the fluid flow generated by point forces at low Reynolds number, and have already been shown to be important in magnetotactic band formation~\cite{guell88} and in many aspects of bacterial behaviour near surfaces~\cite{riedel05,lauga06,hill07}. Hydrodynamic interactions between swimmers have been studied using a variety of theoretical models, including flagella driven micromachines~\cite{nasseri97}, rigidly rotating helices~\cite{kim04}, squirmers~\cite{ishikawa}, linked sphere swimmers~\cite{chris07} and simple `body' and `thruster' models~\cite{cisneros07}. 

A vital concept in understanding the swimming of microscopic organisms is that the Stokes' equations, which govern zero Reynolds number fluid flows, do not possess any intrinsic notion of time. For an incompressible fluid of viscosity $\mu$, the fluid velocity ${\bf u}$ and pressure $p$ satisfy 
\begin{equation}
\mu \nabla^2 {\bf u} - \nabla p = 0 \; , \qquad
\nabla \cdot {\bf u} = 0 \; , 
\label{eq:stokes}
\end{equation}
and the flow throughout the entire fluid is determined by specifying the instantaneous boundary conditions. The fluid moves when the boundaries move and stops when the boundaries stop. If the motion of the boundaries is reversed then the fluid flow is also reversed and each fluid element returns to its original position, a phenomenon known as kinematic reversibility of Stokes flows. This has important consequences for the locomotion of microscopic organisms, for if their motions are reciprocal, such as the opening and closing of a single-hinged scallop~\cite{purcell77}, then kinematic reversibility implies that the forward motion of the first half of the stroke is exactly cancelled during the second half and there is no net motion, a result commonly referred to as the Scallop theorem.

Kinematic reversibility also implies that when the motion of a swimmer ($A$) is reversed, it produces a second swimming stroke ($\bar{A}$) just with the swimmer moving in the opposite direction. We refer to this as the {\em T-dual} swimmer, which may be interpreted as the original swimmer going backwards in time. In this Letter we exploit the time reversal invariance of the Stokes' equations to show that, during any scattering event involving a swimmer and its T-dual, the initial state as $t \rightarrow - \infty$ is recovered exactly in the final state as $t \rightarrow + \infty$, i.e., {\it the angle between the swimmers is unchanged by the scattering}. Experimental verification of the scattering behaviour we describe should be possible using biological or fabricated microswimmers. 

\begin{figure}[b]
\centering
\includegraphics[width=.45\textwidth]{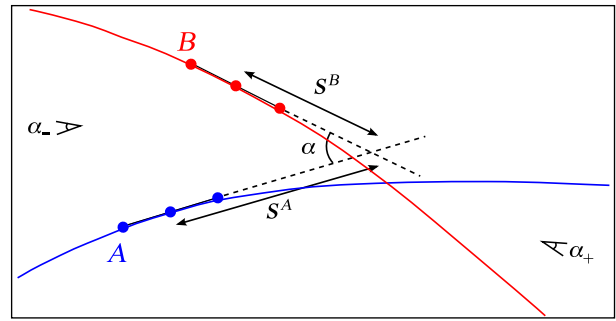}
\caption{(Colour online) Schematic diagram of the geometry of a planar scattering event: see text for details.}
\label{fig:geom}
\end{figure}

For simplicity we focus our attention on planar scattering, however our results naturally generalise to three dimensional geometries. Two swimmers, $A$ and $B$, travel along coplanar trajectories with instantaneous swimming directions ${\bf n}^A(t)$ and ${\bf n}^B(t)$ (see Fig.~\ref{fig:geom}). Provided these directions are not parallel they generate two straight lines which intersect at some point. The angle between these lines defines the {\it angle of incidence}, $\alpha$, and the difference in the distances, $s^A,s^B$, of the two swimmers from the intersection point defines the {\it impact parameter}, $b = s^A - s^B$. The limiting values $(\alpha_{\pm},b_{\pm}) = \lim_{t \rightarrow \pm \infty} (\alpha,b)$ are used to define the initial and final states. The change in orientation of an individual swimmer as a result of the scattering process is described by the {\it scattering angle} 
\begin{equation}
\theta = \arccos \bigl( {\bf n}(t \rightarrow +\infty) \cdot {\bf n}(t \rightarrow -\infty) \bigr) \; , 
\end{equation}
which we take to be positive if the swimmer rotates in an anticlockwise sense and negative otherwise. 

Hydrodynamic scattering may be viewed as providing a map from the initial state $(\alpha_-,b_-)$ to the final state $(\alpha_+,b_+)$. The differences between these two states, defined by the functions $\delta_{\alpha} \coloneq \alpha_+ - \alpha_-$ and $\delta_b \coloneq b_+ - b_-$, describe the tendancy for the swimmers to align ($\delta_{\alpha} < 0$) or cluster ($\delta_b < 0$) via hydrodynamic interactions. Viewing the entire process backwards in time corresponds to the hydrodynamic scattering of the T-dual swimmers ($\bar{B},\bar{A}$) taking the initial state $(\alpha_+,b_+)$ into the final state $(\alpha_-,b_-)$, thereby establishing an isomorphism between the scattering of an arbitrary pair of swimmers and the scattering of their T-duals. In particular the functions $\delta^{(\bar{B},\bar{A})}$ are simply related to the functions $\delta^{(A,B)}$ 
\begin{equation}
\delta^{(\bar{B},\bar{A})} (\alpha_+,b_+) = - \delta^{(A,B)} (\alpha_-,b_-) \; .
\label{eq:tdualfunctions}
\end{equation}
In the case of a pair of mutually T-dual swimmers ($\bar{B}=A;\bar{A}=B$) Eq.~\eqref{eq:tdualfunctions} is sufficient to show that $(\alpha_+,b_+) \equiv (\alpha_-,b_-)$. 

\begin{figure}
\centering
\includegraphics[width=.45\textwidth]{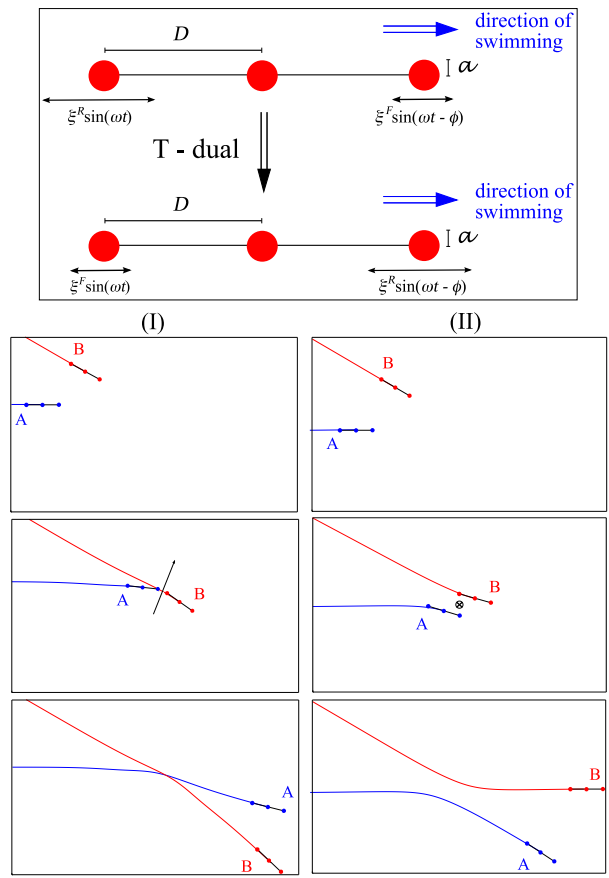}
\caption{(Colour online) Box: Schematic diagram of the Golestanian model swimmer~\cite{ramin} and its T-dual. The swimmer is self T-dual if the two arm amplitudes are equal, $\xi^R = \xi^F$. Below: Exemplary scattering trajectories of two identical, self T-dual Golestanian swimmers obtained using the Oseen tensor description of the hydrodynamics. In (I) the swimmers rotate in the same direction, while in (II) they rotate in opposite directions and exchange trajectories. The rotation axes for the symmetry transformations are indicated in the central panel, which corresponds to the time $t=0$. The initial conditions were $\alpha_- = 30^o$ with $b_- = 3.5D$ for the turn event (I) and $b_- = 2D$ for the exchange event (II).}
\label{fig:gol}
\end{figure}

We motivate this result using symmetry arguments. During any scattering event the quantity $s^A + s^B$ changes from being large and positive to being large and negative. Since it does this continuously it must pass through zero, which we use to define the time $t = 0$. The separation between the two swimmers is given by ${\bf r} = (s^A + s^B) ({\bf n}^A - {\bf n}^B)/2 + (s^A - s^B) ({\bf n}^A + {\bf n}^B)/2$, and is orthogonal to the direction ${\bf n}^A - {\bf n}^B$ at $t = 0$. At this instant reversing the direction of time, followed by a $\pi$ rotation about an axis parallel to ${\bf n}^A - {\bf n}^B$ and passing through the point mid-way between the two swimmers leads to a configuration where $\bar{B},\bar{A}$ have the same positions and orientations as $A,B$, respectively. For mutually T-dual swimmers this returns the same configuration we started with. It follows that $A$'s outgoing trajectory for $t>0$ will be given by $B$'s ingoing trajectory for $t<0$ (with the direction of time reversed) and vice-versa, from which we conclude that the initial and final states are the same. In addition, this construction implies that the swimmers rotate in the same direction and with equal scattering angles, $\theta^A = \theta^B$. We therefore refer to these as {\it turn} events and show an example in Fig.~\ref{fig:gol}(I). 

An exception to this scenario occurs if the swimmers ever become exactly parallel, ${\bf n}^A = {\bf n}^B$. However, taking $t=0$ at this instant and choosing the rotation axis to be parallel to ${\bf n}^A \times {\bf r}$ leads to the same conclusion. This time, since the two swimmers rotate in opposite directions, the constraint that $\alpha_+ = \alpha_-$ can only be met if the scattering angles take the values $\theta = \pm \alpha_-$ independent of $b_-$. In such an event $A$ will rotate so that its outgoing trajectory is parallel to $B$'s ingoing trajectory and vice-versa. We call this an {\it exchange} event, an example of which is shown in Fig.~\ref{fig:gol}(II). Since we expect $\theta \rightarrow 0$ as $b_- \rightarrow \infty$, exchange events can only occur for sufficiently small values of $b_-$. Finally, we comment that, since for purely planar scattering there is no way to cross smoothly between these two cases, they are necessarily separated by some form of discontinuous behaviour. 

There is a subtlety in the foregoing observations. To exactly interchange the swimmers as described, the time $t=0$ must coincide with particular instants during the swimming cycle, otherwise, although the positions and orientations of the swimmers will be the same, the stages they are at during their swimming strokes will not. We have not been able to show generally that the time $t=0$ does indeed coincide with one of these instances (although for this not to be the case would imply rather peculiar properties for the functions $\delta^{(A,\bar{A})}$). However, in numerical tests using the Golestanian model the swimmers do indeed reach $t=0$ at a suitable stage of their stroke. Also, since our discussion has mentioned only the swimming direction, ${\bf n}$, it has been restricted to swimmers that are axisymmetric, requiring only this vector to completely specify their orientation. 

These general symmetry considerations provide a framework for what can be expected in two body swimmer scattering. In the remainder of this Letter we illustrate and extend the results by describing the hydrodynamic scattering of a simple model of linked sphere swimmers first introduced by Najafi and Golestanian~\cite{ramin}. Three spheres, each of radius $a$, are connected by thin rods of natural length $D$, as shown in Fig.~\ref{fig:gol}. By periodically extending and contracting these rods the organism is able to swim in the direction of its long axis. Many different swimming strokes are possible, but we consider here only a simple stroke in which the rods undergo sinusoidal oscillations with amplitudes $\xi^R,\xi^F$ and with a phase lag $\phi$ between them. Viewing this motion backwards in time, one sees that T-duality corresponds to an interchange of the amplitudes, $\xi^R \leftrightarrow \xi^F$. Note that the swimming stroke is unchanged if the amplitudes are equal. This is an example of a special type of swimmer which we refer to as {\em self T-dual}. Swimmers of this type have strokes that `look the same forwards and backwards in time' and yet are not reciprocal: they are time reversal covariant rather than invariant. A number of simple model swimmers commonly referred to in the literature are self T-dual, such as Taylor's rotating torus~\cite{taylor}, Purcell's three link swimmer~\cite{purcell77}, the `pushmepullyou' swimmer~\cite{avron05}, a sinusoidally waving sheet~\cite{taylor} and a rigidly rotating helical filament~\cite{lighthill76}.

The swimming motion of a single Golestanian swimmer may be determined analytically using the Oseen tensor to describe the hydrodynamics~\cite{ramin}. This approach may also be applied to determine the interactions between two swimmers~\cite{chris07}. These interactions prescribe how the positions and orientations of each swimmer are altered by the fluid flow generated by the other during one swimming stroke. By iterating theses changes numerically we are able to generate the trajectories of the two swimmers during a scattering event. Since this approach is based on the Oseen tensor approximation to the hydrodynamics it is only valid so long as the separation $r$ between the swimmers is large compared to the size of the spheres, i.e., $a/r \ll 1$. An important feature of the interactions is that they are strongly sensitive to the relative phase $\eta$ of the two swimmers, which therefore has a significant influence on the type of behaviour that is observed~\cite{chris07}. 
 
\begin{figure}
\centering
\includegraphics[width=.45\textwidth]{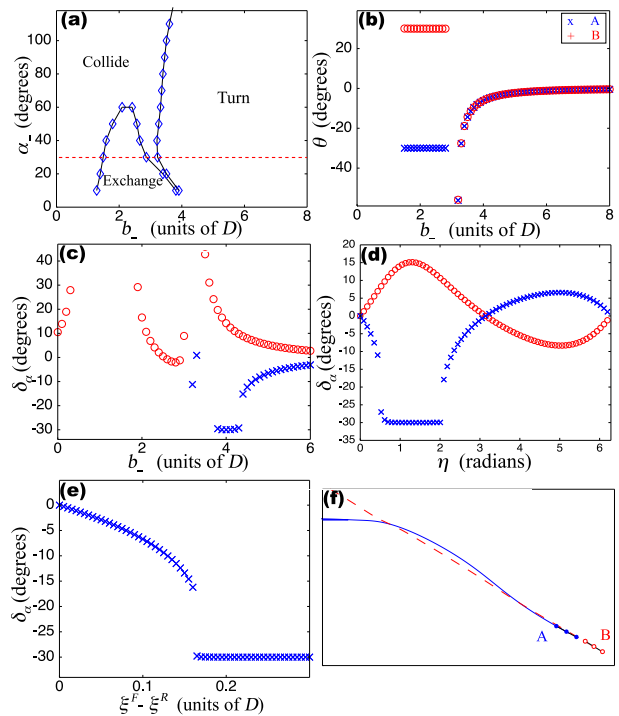}
\caption{(Colour online) Hydrodynamic scattering of two identical Golestanian swimmers obtained using the Oseen tensor description of the hydrodynamics. (a) The type of scattering observed for different values of the initial conditions $\alpha_-,b_-$ when the swimmers are in phase, $\eta = 0$. (b) Dependence of the scattering angle on $b_-$ for a fixed value of $\alpha_- = 30^o$, corresponding to the dashed line in (a). The change in alignment $\delta_{\alpha}$ is shown as a function of (c) $b_-$ for $\alpha_- = 30^o, \eta = \pi /2$ and (d) $\eta$ for $\alpha_- = 30^o, b_- = 4D$, for two extensile (circles) and two contractile (crosses) swimmers. (e) Variation of $\delta_{\alpha}$ with the difference in amplitudes $\xi^F - \xi^R$ for two identical swimmers with $\alpha_- = 30^o, b_- = 4D$ and $\eta = \pi /2$ and (f) exemplary trajectory of the bound state formation for $\xi^F - \xi^R = 0.17D$.}
\label{fig:align}
\end{figure}

We consider first two identical swimmers that are in phase, $\eta = 0$, and have equal arm amplitudes, $\xi^R = \xi^F$, for which the swimming stroke is self T-dual. For all trajectories $(\alpha_+,b_+) = (\alpha_-,b_-)$ in accordance with the symmetry arguments we have presented above. The type of scattering event (exchange or turn) that occurs is shown as a function of the two initial conditions $\alpha_-,b_-$ in Fig.~\ref{fig:align}(a), together with a detailed cut showing how the scattering angle $\theta$ varies with $b_-$ for a fixed value of $\alpha_- = 30^o$ (Fig.~\ref{fig:align}(b)). There is a wide range of initial conditions at small values of $b_-$ for which the scattering is of the exchange type and $\theta = \pm \alpha_-$. At larger values of $b_-$ the scattering is always of the turn type with the scattering angle decaying to zero as $b_- \rightarrow \infty$. In the region labelled Collide the swimmers approach so closely that the Oseen tensor description of the hydrodynamics is no longer valid and we are unable to determine what happens during the scattering. In our simulations we took this to occur when the minimum separation between any two spheres became less than $10a$. 

We now outline how the properties of swimmer scattering change when the two swimmers are not mutually T-dual and our preceeding symmetry arguments no longer apply. In Fig.~\ref{fig:align}(c) we show the dependence of $\delta_{\alpha}$ on $b_-$ for $\alpha_- = 30^o$ and $\eta = \pi /2$. For a pair of identical extensile swimmers with $(\xi^R,\xi^F) = (0.3D,0.1D)$ the scattering predominantly leads to an increase in the angle between the two swimmers, i.e., $\delta_{\alpha} > 0$. By contrast, a pair of identical contractile swimmers with $(\xi^R,\xi^F) = (0.1D,0.3D)$ predominantly shows hydrodynamically induced alignment, $\delta_{\alpha} < 0$. The relative phase is important for these results, as illustrated in Fig.~\ref{fig:align}(d). In particular, the sign of $\delta_{\alpha}$ is different for $0 < \eta < \pi$ and $\pi < \eta < 2\pi$ for both extensile and contractile swimmers, revealing that there is not a simple, direct link between the type of swimming stroke and a tendency for hydrodynamic alignment. Moreover, for $\eta = 0,\pi$ we find the intriguing result that $\delta_{\alpha} = 0$ even though the swimmers are not mutually T-dual, a result which is also found for all other initial conditions $(\alpha_-,b_-)$ so long as the swimmers do not collide (not shown). 

One further feature of the scattering of contractile swimmers deserves mention, namely the appearance in Figs.~\ref{fig:align}(c) and (d) of a range of scattering events for which $\delta_{\alpha}$ takes the constant value $-30^o$, corresponding to $\alpha_+ = 0^o$. This represents the formation of a bound state in which the two swimmers are exactly aligned one behind the other. An example of swimmer trajectories during the formation of this bound state is shown in Fig.~\ref{fig:align}(e). On either side of the bound state region $\delta_{\alpha}$ takes a value somewhat larger than $-30^o$, indicating that $\delta_{\alpha}$ is discontinous at the transition from scattering to bound state formation. This qualitative observation is supported by the behaviour when the transition is approached from a different direction, that of increasing the difference in amplitudes $\xi^F - \xi^R$ at a fixed value of $b_- = 4D$ as shown in Fig.~\ref{fig:align}(d). We vary this difference in amplitudes whilst holding the product $\xi^R\xi^F$ fixed to maintain an approximately constant swimming speed~\cite{ramin,chris07}. As $\xi^F - \xi^R$ is increased from zero $\delta_{\alpha}$ decreases smoothly until it reaches the value $0.16D$ where there is an abrupt transition to the bound state, accompanied by a substantial discontinuity in $\delta_{\alpha}$.\\[1mm]

We have described the constraints imposed on the hydrodynamic scattering of two swimmers by the time reversal invariance of the Stokes' equations. The most striking observation concerns T-dual swimmers, which have strokes that map onto each other under time reversal: for scattering events involving two such swimmers the angle and impact parameter between their trajectories are the same before and after the collision. For swimmers unrelated by T-duality we show numerically that scattering is complex, with the possibility of changes in the angle between the two swimmers of several tens of degrees or the formation of bound states. Experiments on biological or fabricated microswimmers should show these striking differences between pairs of mutually T-dual and symmetry unrelated swimmers. In future work it may be possible to extend our ideas to more than two swimmers, thereby constraining the properties expected of large groups or swarms, and providing a connection to continuum models. 

We are grateful to Mike Cates, Scott Edwards, Davide Marenduzzo, 
and Vic Putz for enlightening discussions about this work.

\end{document}